\documentclass[longnamesfirst,apjl]{emulateapj}
\usepackage{apjfonts,natbib,epsfig}
\tighten

\newcommand{\beq}{\begin{equation}}
\newcommand{\eeq}{\end{equation}}
\newcommand{\bea}{\begin{eqnarray}}
\newcommand{\eea}{\end{eqnarray}}

\begin{document}
\title{Cluster Merger Shock Constraints on Particle Acceleration and Nonthermal Pressure in the Intracluster Medium}
\author{
Ehud Nakar\altaffilmark{1},
Milo\v s Milosavljevi\'c\altaffilmark{2}, and
Daisuke Nagai\altaffilmark{1}} \altaffiltext{1}{Theoretical
Astrophysics, Mail Code 130-33, California Institute of Technology,
1200 East California Boulevard, Pasadena, CA 91125.}
\altaffiltext{2}{Department of Astronomy, University of Texas, 1
University Station C1400, Austin, TX 78712.}

\righthead{SHOCK CONSTRAINTS ON NONTHERMAL PRESSURE IN CLUSTERS}

\lefthead{NAKAR ET AL.}

\begin{abstract}

X-ray observations of galaxy cluster merger shocks can be used to
constrain nonthermal processes in the intracluster medium (ICM). The
presence of nonthermal pressure components in the ICM, as well as the
shock acceleration of particles and their escape, all affect shock
jump conditions in distinct ways. Therefore, these processes can be
constrained using X-ray surface brightness and temperature maps of
merger shock fronts. Here we use these observations to place
constraints on particle acceleration efficiency in intermediate Mach
number (${\cal M} \approx 2-3$) shocks and explore the potential to
constrain the contribution of nonthermal components (e.g., cosmic
rays, magnetic field, and turbulence) to ICM pressure in cluster
outskirts.  We model the hydrodynamic jump conditions in merger shocks
discovered in the galaxy clusters A520 (${\cal M} \approx 2$) and 1E
0657$-$56 (${\cal M} \approx 3$) using a multifluid model comprised of
a thermal plasma, a nonthermal plasma, and a magnetic field.  Based on
the published X-ray spectroscopic data alone, we find that the
fractional contribution of cosmic rays accelerated in these shocks is
$\lesssim 10\%$ of the shock downstream pressure. Current observations
do not constrain the fractional contribution of nonthermal components
to the pressure of the undisturbed shock upstream. Future X-ray
observations, however, have the potential to either detect particle
acceleration in these shocks through its effect on the shock dynamics,
or to place a lower limit on the nonthermal pressure contributions in
the undisturbed ICM. We briefly discuss implications for models of
particle acceleration in collisionless shocks and the estimates of
galaxy cluster masses derived from X-ray and Sunyaev-Zel'dovich effect
observations.


\end{abstract}

\section{Introduction}

\setcounter{footnote}{0}

Astrophysical collisionless shocks are the likely sources of the
observed extra-solar high energy cosmic rays
\citep[e.g.,][]{Bell:78,Blandford:78}. Continuous observational
effort has not yet yielded direct evidence for acceleration of
cosmic ray nuclei in collisionless shocks, although recently,
tentative indirect evidence for such acceleration was identified in
the morphology of the Tycho supernova remnant \citep{Warren:05}, and
in the high-energy gamma-ray emission near the RX J1713.7$-$3946
remnant \citep{Aharonian:07}. The theory of diffusive shock
acceleration \citep[e.g.,][and references therein]{Blandford:87}
predicts that the acceleration efficiency and the spectrum of
accelerated particles depend on the Mach number and other parameters
of the shock \citep[e.g.,][and references therein]{Giacalone:97}.
The presence of fossil cosmic rays in the pre-shock medium, e.g.,
from previous shocks, can also affect the acceleration efficiency
\citep[e.g.,][and references therein]{Kang:07a,Kang:07b}. We here
argue that merging galaxy clusters are laboratories in which
theories of cosmic ray acceleration in intermediate Mach number
shocks can be tested. Observational constraints on particle
acceleration in such shocks are especially interesting as numerical
simulations suggest that these are the source of a large fraction of
the cosmic rays accelerated in galaxy clusters
\citep[e.g.,][]{Ryu:03}.

Fossil cosmic rays,\footnote{Potential acceleration sites of fossil
cosmic rays are past accretion and merger shocks, giant radio
sources, supernovae, and turbulence in the ICM
\citep[e.g.,][]{Berezinsky:97}} magnetic field, and turbulence may
all contribute to the pressure of the intracluster medium (ICM),
thereby modifying its hydrodynamic behavior. Such contribution may
alter the interpretation of observations that ignore them. For
example X-ray observations have recently been used to estimate shock
velocities in two merging clusters (see below) neglecting nonthermal
components; improved estimates of shock velocities in these and
other clusters may have to take into account the cosmic ray and
other nonthermal contributions to the ICM pressure. Nonthermal
pressure is also a source of systematic bias when cluster masses are
estimated from X-ray and Sunyaev-Zel'dovich effect (SZE)
measurements that are made assuming an hydrostatic equilibrium
between gravitational forces and thermal pressure gradients in the
ICM \citep[e.g.,][and references
therein]{Ensslin:97,Rasia:06,Nagai:07a}.  These nonthermal biases
limit the effectiveness of upcoming cluster surveys in the quest to
place constraints on the expansion history of the universe.

Evidence for the nonthermal activity in clusters is growing.
Observed radio and hard X-ray emissions in clusters suggest a
presence of relativistic electrons with Lorentz factor of $\sim
10^4$. This also suggests a presence of relativistic protons that
could have been accelerated by the same mechanism that has
accelerated the electrons. Direct evidence for cosmic ray ions in
the ICM is, however, still lacking.  The nondetection in {\it EGRET}
data of gamma-ray emission expected from neutral pion decay in
cosmic ray collisions in the ICM \citep[e.g.,][]{Reimer:03} has so
far placed upper limits on the fraction of cosmic ray pressure to
$\lesssim 20\%-30\%$ in several nearby rich clusters
\citep{Ensslin:97,Pfrommer:04}.  Most cluster atmospheres are also
substantially magnetized, with typical field strengths of order a
few $\mu$G out to Mpc radii \citep[][and references
therein]{Carilli:02,Govoni:04,Govoni:06}. There is likely to be
considerable variation in field strengths ($\sim 0.1-40
\mu\textrm{G}$) and topologies within clusters. Thus while magnetic
fields are likely to provide a significant contribution to the
pressure in some regions \citep[e.g., along some cold fronts;
see][]{Vikhlinin:01}, it is yet unclear what is the average
energetics in the cluster outskirts.  Numerical simulations of
cluster formation also suggest that subsonic gas motions
(turbulence) contribute substantial nonthermal pressure in clusters
\citep[e.g.,][]{Norman:99,Ricker:01,Nagai:03,Faltenbacher:05,
Dolag:05,Rasia:06,Nagai:07a}.  Further investigations of nonthermal
phenomena in clusters are hence critical for the success of upcoming
X-ray and SZE cluster surveys, as our ability to estimate cluster
masses hinges on a precise characterization of the nonthermal
components.

In this work, we show that shock waves that form during merging of
galaxy clusters can provide unique constraints on nonthermal processes
in clusters.  Recent {\it Chandra} X-ray observations have revealed
that shock waves with Mach numbers ${\cal M} \approx 2-3$ that form
during the merging of galaxy clusters are accompanied by distinct
X-ray surface brightness and temperature discontinuities.  To date, a
density and temperature jump along axis of symmetry of a cluster
merger bow shock has been recovered in the cluster A520
\citep{Markevitch:05} and in 1E 0657$-$56
\citep{Markevitch:02,Markevitch:06}. The relative magnitudes of the
density and temperature jump depend on the contribution of nonthermal
components (e.g., cosmic rays, magnetic field, turbulence) to the
pressure of the undisturbed, upstream ICM.  They also depend on the
efficiency of particle acceleration in the shock, on the escape of the
accelerated particles from the shock, and on the amplification of
turbulence in the shock.

Here we make the first attempt to use the X-ray observations of
merging clusters to constrain their particle acceleration and the
fractional contribution of nonthermal components to the pressure
budget of the ICM. In \S~\ref{sec:theory}, we model the hydrodynamic
jump conditions and Mach number using the multi-fluid approximation.
We model the effect of cosmic rays, a tangled magnetic field, and
turbulence, on observed gas jump conditions in the shock. In
\S~\ref{sec:results}, we utilize this model to derive constraints on
nonthermal components in merger shocks in galaxy clusters A520 and 1E
0657$-$56. We show that current observations can limit the efficiency
of particle acceleration, and that the fractional contribution of
nonthermal fluids to ICM pressure may be constrained with future,
improved, X-ray spectroscopy and SZE observations. In
\S~\ref{sec:discussion}, we discuss implications of these results in
context of particle acceleration models and for the estimates of
galaxy clusters masses with X-ray and SZE observations. In
\S~\ref{sec:conclusions}, we summarize our main conclusions.

\section{Limits on Particle Acceleration in Merger Shocks and Nonthermal Pressure in ICM}
\label{sec:theory}

\subsection{Shock Jump Conditions}

The ICM fluid consists of a thermal component (electrons and ions in
mutual thermal equilibrium) and number of nonthermal components
(nonthermal cosmic rays and magnetic fields comoving with the
thermal component).  Turbulent gas motions and electromagnetic waves
can also contribute pressure to the ICM. We ignore the
electromagnetic waves in the current treatment. We also temporarily
ignore turbulence, but return to discuss its role on qualitative
level in \S~\ref{sec:turbulence}.  The ICM fluid can be modeled with
an adiabatic equation of state with an effective adiabatic index
$\gamma$.\footnote{We define the adiabatic indices via
$\gamma=1+P/\varepsilon$, where $\varepsilon$ is the internal energy
density. The adiabatic index thus calculated may differ from
$\partial \ln P/\partial \ln\rho$.}  In what follows, we use indices
``u'' and ``d'' to denote the shock upstream and downstream,
respectively. Assuming cylindrical symmetry, conservation of
density, momentum, and energy fluxes across the shock dictates jump
conditions on the symmetry axis of the bow shock that read
\begin{eqnarray} \label{EQ:shock_jump}
  \rho_{\rm u} v_{\rm u}&=&\rho_{\rm d} v_{\rm d} , \nonumber\\
  P_{\rm u}+\rho_{\rm u}v_{\rm u}^2&=&P_{\rm d}+\rho_{\rm d}v_{\rm d}^2 , \nonumber\\
  v_{\rm u}\left(\frac{1}{2}\rho_{\rm u}v_{\rm u}^2+\frac{\gamma_{\rm u}}{\gamma_{\rm u}-1}P_{\rm u}\right)&=&v_{\rm d}\left(\frac{1}{2}\rho_{\rm d}v_{\rm d}^2+\frac{\gamma_{\rm d}}{\gamma_{\rm d}-1}P_{\rm d}\right) ,
\end{eqnarray}
where $\rho_i$ is the mass density of the thermal gas (the mass
density of the nonthermal particles is negligible), $v_i$ is the
fluid velocity, and $P_i$ is the total pressure  ($i=\{u,d\}$)

The pressure is a sum of electronic, ionic, cosmic ray, and magnetic
field contributions, $P_i=P_{{\rm e},i}+P_{{\rm i},i}+P_{{\rm
CR},i}+P_{B,i}$.  The effective adiabatic index $\gamma_i$ equals
\begin{eqnarray}
\label{EQ:gamma}
    \frac{\gamma_i}{\gamma_i-1}&=&\frac{1-\epsilon_{{\rm nt},i}}{2}
    \left(\frac{\gamma_{{\rm i},i}}{\gamma_{{\rm i},i}-1}
    +\frac{\gamma_{{\rm e},i}}{\gamma_{{\rm e},i}-1}\right)\nonumber\\& &
    +\epsilon_{{\rm CR},i}\frac{\gamma_{\rm CR}}{\gamma_{\rm CR}-1}
    +\epsilon_{B,i}\frac{\gamma_{B,i}}{\gamma_{B,i}-1}.
\end{eqnarray}
where $\epsilon_{{\rm CR},i}\equiv P_{{\rm CR},i}/P_i$ and
$\epsilon_{B,i}\equiv P_{B,i}/P_i$ are, respectively, the fractional
contribution of cosmic rays and magnetic fields to the total pressure,
while $\epsilon_{{\rm nt},i}=\epsilon_{{\rm CR},i}+\epsilon_{B,i}$ is
the total fractional contribution of nonthermal pressure
components. The temperature jump across the shock of the thermal
electrons is then
\begin{equation}\label{EQ:Tjump}
\tau\equiv \frac{T_{\rm e,d}}{T_{\rm e,u}}=\frac{(1-\epsilon_{\rm
nt,d})[2\gamma_{\rm u}/(\gamma_{\rm
u}-1)-r^{-1}-1]}{(1-\epsilon_{\rm nt,u})[2\gamma_{\rm
d}/(\gamma_{\rm d}-1)-r-1]},
\end{equation}
where $r\equiv \rho_{\rm d}/\rho_{\rm u}$ is the compression ratio.

\subsection{Thermal Electron Relativistic Corrections}

Thermal ions are nonrelativistic in cluster shocks and thus
$\gamma_{{\rm i},i}=5/3$.  The adiabatic index of thermal electrons
may differ from this value because of relativistic corrections. The
adiabatic index of thermal electrons equals the ratio of the
rest-frame pressure to internal energy density, $\gamma_{\rm
e}=1+P_{\rm e}/ \varepsilon_{\rm e}$, where $\varepsilon_{\rm e}$ is
the internal electron energy density (e.g., \citealt{Achterberg:84})
\begin{equation}
\label{eq:gamma_e} \gamma_{\rm e} = 1 + \frac{1}{3}
\frac{\int_0^\infty 4\pi p^4 (p^2/c^2+m_{\rm e}^2)^{-1/2}f_{\rm
e}(p) dp}{\int_0^\infty 4\pi p^2 [(p^2c^2+m_{\rm
e}^2c^4)^{1/2}-m_{\rm e}c^2]f_{\rm e}(p) dp} ,
\end{equation}
where
\begin{equation}  f_{\rm e}(p) \propto \exp\left[-\frac{(p^2c^2+m_{\rm
e}^2c^4)^{1/2}}{kT_{\rm e}}\right]
\end{equation}
is the thermal electron momentum distribution. Defining the electron
temperature in units of the electron rest energy $\Theta_{\rm
e}\equiv kT_{\rm e}/ m_{\rm e}c^2$, the adiabatic index in equation
(\ref{eq:gamma_e}) can conveniently be expressed in terms of
modified Bessel functions of the second kind (e.g.,
\citealt{Kunik:03})
\begin{equation}
\label{eq:gamma_e_bessel} \gamma_{{\rm e},i}=1+\Theta_{{\rm e},i}
\left[3\Theta_{{\rm e},i}+ \frac{K_1(\Theta_{{\rm
e},i}^{-1})}{K_2(\Theta_{{\rm e},i}^{-1})}-1\right]^{-1}
\end{equation}
In the limit $\Theta_{\rm e}\ll 1$, relevant to cluster merger shocks,
the adiabatic index is $\gamma_{\rm e}= 5/3-(5/6) \Theta_{\rm e}+{\cal
O}(\Theta_{\rm e}^2)$. Note that when $kT_{\rm e,d}$ or $kT_{\rm e,u}$
becomes comparable to $m_{\rm e}c^2$, the right side of equation
(\ref{EQ:Tjump}) depends on the electron temperatures.  For example,
if the nonthermal pressures are ignored (i.e., $\epsilon_{{\rm
nt},i}=0$), a relativistic correction of about $10\%$ applies to the
expected temperature jump and inferred Mach number for conditions
similar to those in the bow shock in 1E 0657$-$56 ($T_{\rm
u}=10\textrm{ keV}$ and $r=3$), while a correction of only about $1\%$
applies for the lower-temperature and weaker shock in A520 ($T_{\rm
u}=5\textrm{ keV}$ and $r=2.3$).

\subsection{Fossil Cosmic Rays}

The undisturbed ICM may contain an intracluster population of fossil
cosmic rays that could have been produced in the high Mach number
accretion shocks (${\cal M}\sim10-100$), in previous merger shocks, in
active galactic nuclei, and in starburst-associated phenomena
\citep[see,
e.g.,][]{Berezinsky:97,Fujita:01,Miniati:01,Gabici:03,Sarazin:04}.
After a merger bow shock passes a fluid element in the ICM, the fluid
element will contain the original fossil cosmic rays, some of which
may have been further accelerated in the shock. The shock may also
accelerate new cosmic rays drawn from the thermal pool and accelerated
in the shock for the first time.

The cosmic ray adiabatic index depends on whether their pressure is
dominated by Newtonian or relativistic particles, which in turn
depends on the details of the cosmic ray spectrum
\citep[e.g.,][]{Achterberg:84}.  The effective cosmic ray adiabatic
index lies in the range $4/3 \leq \gamma_{\rm CR} \leq 5/3$; the
electron cosmic rays are mildly or fully relativistic while the
protons that dominate the cosmic rays pressure can be Newtonian or
relativistic. Thus, assuming that electron and proton cosmic rays
are close to mutual equipartition, we expect $\gamma_{\rm CR}$ to
range between $\gamma_{\rm CR}\approx 4/3$ (for relativistic
electrons and protons) and $\gamma_{\rm CR} \approx 13/9$
\citep[relativistic electrons and Newtonian protons;
e.g.,][]{Konigl:80}. While in principle $\gamma_{\rm CR}$ may differ
between the upstream and the downstream, when there is a significant
population of upstream cosmic rays, it is reasonable to expect that
these will also dominate the downstream, and thus that $\gamma_{\rm
CR}$ is approximately constant across the shock.

In addition to being reaccelerated by shock acceleration mechanism,
fossil cosmic rays may be accelerated adiabatically during the shock
compression of the magnetic field. However, shock compression may
very well be nonadiabatic as the Larmor radius of a mildly
relativistic electron in a microgauss field, $r_{\rm L}\sim
10^9\textrm{ cm}$, may be much larger than the transition layer over
which the fluid density jumps.\footnote{If pressure across the
transition layer is not mediated by cosmic rays but by plasma
instabilities, the width of the transition layer will be of the
order of the plasma skin depth of the thermal protons, which is
$\delta \sim 10^7\textrm{ cm}$. On the other hand, adiabatic
acceleration will take place when $r_{\rm L}\ll \delta$.} Therefore,
it is possible that cosmic rays do not gain energy during
compression. Nevertheless, the compression increases the cosmic ray
number density and therefore its pressure by at least a factor of
$r$. Thus we express the fractional cosmic ray pressure in the
downstream as
\begin{equation} \label{eq:nofossil} \epsilon_{\rm
CR,d}=r\left(\frac{P_{\rm u}}{P_{\rm d}}\right)\epsilon_{\rm
CR,u}+\epsilon_{\rm acc}
\end{equation}
where the first term is the pressure of the fossil cosmic rays if
the energy of individual cosmic rays remains unchanged as they pass
through the shock, and the second term is any additional pressure of
cosmic rays that are accelerated in the shock. The acceleration term
includes pressure due to accelerated particles from the upstream
thermal pool that are added to the nonthermal pool, and also due to
fossil nonthermal particles that are reaccelerated by a shock
acceleration process and/or by adiabatic compression.

\subsection{The Cluster Magnetic Field}
\label{sec:magnetic_field}

The effect of the magnetic field on the shock depends on its
strength and topology, both of which are constrained very poorly,
especially in the cluster outskirts. Faraday rotation measurements
suggest that the magnetic pressure can reach $\sim 10\%$ of the
thermal pressure in some clusters or regions within clusters, and
that in others it is less than $1\%$
\citep[e.g.,][]{Carilli:02,Govoni:04,Govoni:06}.\footnote{A magnetic
field that is tangled on very small scales is not well constrained
by the Faraday rotation measurements.} Given the large uncertainty
in the field strength, we treat $\epsilon_B$ as a free parameter in
what follows.

We assume that the magnetic field of the undisturbed ICM is tangled
and isotropic on scales relevant to the hydrodynamics of the cluster
merger shock. This approximation allows us to use the unmagnetized
form of the jump conditions, equation (\ref{EQ:shock_jump}), to
describe the effect of magnetic field on shock hydrodynamics. The
approximate coherence length of the field in some galaxy clusters,
$\sim 10\textrm{ kpc}$, is shorter than the curvature scale of the
merger shock, $\sim 100\textrm{ kpc}$, so that one can average over
the fluctuating field orientation near the shock.\footnote{The same
approximation was employed by \citet{Markevitch:05}.} Furthermore,
the measurements of density and temperature jumps across the shock
are based on deprojection of the X-ray map assuming a cylindrical
symmetry of a bow shock---thus an averaging of observables on the
shock curvature scale is implicit in the reported shock jump data.

Even if the upstream field is isotropic, the downstream field may be
anisotropic as a result of a preferential amplification of the
magnetic field component that is perpendicular to the shock,
$B_\perp$. In this case the perpendicular component will still be
isotropic in the plane of the shock, and can thus be parameterized
by the ratio of the average parallel and perpendicular field energy
densities $b\equiv \langle B_\perp^2\rangle/2\langle
B_\parallel^2\rangle$. The effective adiabatic index of the magnetic
fluid, $\gamma_B$, which is defined as the parameter that correctly
quantifies the behavior of the magnetic pressure in equation
(\ref{EQ:gamma}) and may depend on the orientation of the magnetic
field, can be expressed in terms of the averages of the components
${\cal T}^{\rm em}_{\mu\nu}$ of the electromagnetic part of the
energy-momentum tensor in the shock frame via
\begin{equation}
\frac{\gamma_B}{\gamma_B-1}=\frac{\langle {\cal T}^{\rm em}_{10}\rangle}{\beta \langle {\cal T}^{\rm em}_{11}\rangle} .
\end{equation}
Here, $\beta$ is the shock velocity in units of the speed of light,
and $\Gamma\equiv (1-\beta^2)^{-1/2}$ is the Lorentz factor of the
shock.  Components of the energy-momentum tensor read
\begin{eqnarray}
{\cal T}^{\rm em}_{11}&=&\frac{1}{8\pi}[(2\Gamma^2-1)B_\perp^2-B_\parallel^2] ,\nonumber\\
{\cal T}^{\rm em}_{10}&=&\frac{1}{4\pi}  \Gamma^2\beta B_\perp^2 .
\end{eqnarray}
With these, for a nonrelativistic shock we obtain
\begin{equation}\label{eq:gammaB}
\gamma_B=\frac{4b}{2b+1} .
\end{equation}
The value of $\gamma_B$ in equation (\ref{eq:gammaB}) equals
$\gamma_B=2$ in a perpendicular shock ($b=\infty$) and  vanishes in a
parallel shock ($b=0$). If the field is isotropic on average in the
shock upstream ($b=1$), we recover the value $\gamma_{B,{\rm
u}}=4/3$. For an isotropic upstream field, the downstream field will
have $b \geq 1$ and thus $4/3 \leq \gamma_{B,{\rm d}}\leq 2$.

Relating $\epsilon_{B,{\rm d}}$ to $\epsilon_{B,{\rm u}}$ requires
an understanding of the shock structure. But even if the shock
structure is unknown, we can limit $\epsilon_{B,{\rm d}}$ in two
extreme cases, when field generation in the shock itself (as may
take place in unmagnetized shocks) is neglected. Assuming that
magnetohydrodynamic jump conditions apply within the shock
transition, and that the parallel and the perpendicular fields do
not transform into each other, then $B_{\|,{\rm d}}=B_{\|,{\rm u}}$
and $B_{\bot,{\rm d}}=rB_{\bot,{\rm u}}$, implying that
$\epsilon_{B,{\rm d}}=(2r^2-1)(P_{\rm u}/P_{\rm d})\epsilon_{B,{\rm
u}}$. The other extreme assumption, which was previously made by
\citet{Markevitch:05}, is that the field is isotropic and remains
isotropic throughout the shock compression, in which case
$\epsilon_{B,{\rm d}}=r^{4/3}(P_{\rm u}/P_{\rm d})\epsilon_{B,{\rm
u}}$, and thus the magnetic field behaves as a relativistic or
photon gas that is adiabatically compressed in the shock. We expect
$\epsilon_{B,{\rm d}}$ to lie between these two limits, assuming
that no new field is generated within the shock.

\subsection{Escape of the Highest Energy Cosmic Rays}
\label{sec:escape}

Another process that may affect the shock and that we have ignored so
far is the escape of the highest-energy cosmic rays that are
accelerated in the shock. This process may take place if the shock
efficiently produces a hard cosmic ray spectrum, and the
highest-energy cosmic rays can escape and remove energy from the shock
transition \citep[see, e.g.,][]{Achterberg:84}. This energy leakage
can be parameterized by the fraction $Q$ of the incoming energy flux that
escapes the shock. The leakage affects the shock jump conservation
relations such that the left-hand side of the energy flux conservation
relation---the last of equations (\ref{EQ:shock_jump})---must be
multiplied by the factor $1-Q$. With this, equation (\ref{EQ:Tjump})
generalizes into
\begin{equation}\label{EQ:Tjump_gen}
\tau=\frac{(1-\epsilon_{\rm
nt,d})[2(1-Q)\gamma_{\rm u}/(\gamma_{\rm
u}-1)-r^{-1}-1+Qr/(r-1)]}{(1-\epsilon_{\rm nt,u})[2\gamma_{\rm
d}/(\gamma_{\rm d}-1)-r-1+Qr^2/(r-1)]} .
\end{equation}
The effect of escape is similar to the effect of particle
acceleration, as it removes a fraction of the incoming energy from
the downstream thermal fluid, thereby reducing the temperature jump.
Note, however, that if the measured value of $\tau$ is below the one
expected in a thermal shock, the deviation from the thermal
temperature jump can only partially be attributed to the escape,
since when $\epsilon_{\rm acc}=0$, the value of $Q$ must by
definition be zero.

\subsection{Sensitivity of the Temperature Jump}

We have one relation (equation \ref{EQ:Tjump} or its generalized
form, equation \ref{EQ:Tjump_gen}) and several unknown parameters
describing the upstream and downstream contribution of the cosmic
rays and magnetic field to the fluid pressure. Without making
assumptions about the nature of the nonthermal fluids, given a
measurement of the density jump $r$ and the temperature jump $\tau$,
we can place constraints in the joint parameter space spanned by
these parameters, but cannot recover the fractional pressures
themselves. Additionally, when only soft X-ray data are available
(as from {\it Chandra} and {\it XMM-Newton}), the uncertainties in
$\tau$ typically exceed the uncertainties in $r$. Therefore, before
we proceed to explore the joint parameter space, we discuss the
effect of the variation of various parameters on $\tau$, assuming
that $r$ is accurately measured and is held fixed. The temperature
jump is most sensitive to the fractional pressure in cosmic rays
accelerated or reaccelerated in the shock, $\epsilon_{\rm acc}$. A
nonzero value of $\epsilon_{\rm acc}$ reduces the value of $\tau$,
e.g., in the ideal case in which the upstream nonthermal pressure
vanishes, $\epsilon_{\rm nt,u}=0$, we find that $\tau$ drops by
about a factor of two between $\epsilon_{\rm acc}=0$ and
$\epsilon_{\rm acc}=0.3$ ($r=2-3$). The reason for this sensitive
dependence is that the production of accelerated particles in the
shock saps a fraction of the incoming energy flux out of the thermal
component of the downstream, thereby reducing $\tau$. Increasing $Q$
results in a similar effect, for similar reason, on $\tau$.

The effect of changing $\epsilon_{\rm CR,u}$ and $\epsilon_{B,{\rm
u}}$ on the temperature jump is more subtle since a high nonthermal
pressure in the shock upstream implies that there will also be a
high nonthermal pressure in the downstream; it is the balance
between the two that determines $\tau$. Therefore, $\tau$ depends
only weakly on the upstream nonthermal components. Increasing
$\epsilon_{\rm CR,u}$ while keeping the rest of the parameters
constant always increases $\tau$ slowly. For example assuming that
the magnetic pressure vanishes ($\epsilon_{B,{\rm u}}=0$) and that
$\gamma_{\rm CR}=4/3$, $\tau$ increases by a factor of $\approx 1.2$
between $\epsilon_{\rm CR,u}=0$ and $\epsilon_{\rm CR,u}=0.3$ for
$r=2.3$ and $\epsilon_{\rm acc}=0$. Having a constant $\epsilon_{\rm
acc}=0.3$ reduces the change in $\tau$ to a factor of $\approx 1.1$.
Increasing $\epsilon_{B,{\rm u}}$ also mostly results in a slightly
larger temperature jump. If the perpendicular field is strongly
amplified in the shock, $\gamma_B$ jumps in the shock, thereby
increasing the effective downstream adiabatic index $\gamma_{\rm
d}$, and thus increasing $\tau$. If on the other hand we treat the
field as a relativistic gas, then $\epsilon_{B,{\rm u}}$ and $\tau$
are positively correlated for $\epsilon_{\rm acc}=0$ and are weakly
anticorrelated for $\epsilon_{\rm acc} \approx 0.15$.

In conclusion, $\tau$ is strongly anticorrelated with $\epsilon_{\rm
acc}$ and is typically weakly positively correlated with
$\epsilon_{\rm CR,u}$ and $\epsilon_{B,{\rm u}}$. Therefore an
accurate measurement of $\tau$ and $r$ can tightly constrain
$\epsilon_{\rm acc}$. If the measured value of $\tau$ falls below
that expected in a purely thermal hydrodynamic shock given a precise
measurement of $r$, then a nontrivial lower and upper bound can be
placed on $\epsilon_{\rm acc}$, but such a constraint cannot be
obtained for $\epsilon_{\rm CR,u}$ and $\epsilon_{B,{\rm u}}$. If,
on the other hand, the measured value of $\tau$ is higher than that
expected in a purely thermal hydrodynamic shock, only an upper limit
on $\epsilon_{\rm acc}$ can be placed, i.e., $\epsilon_{\rm acc}=0$
remains viable.  Then, however, the measurement places a lower limit
on the upstream nonthermal pressure. However, based on the
measurement of $r$ and $\tau$ alone, one cannot separate the partial
contributions of the cosmic ray and magnetic components.

\subsection{Implications for Shock Velocity Estimation}
\label{sec:mach_number}

The velocity of cluster merger shock is of great interest since it
is a stepping stone toward relating the dynamics of the ICM to the
dynamics of the dark matter in galaxy cluster mergers
\citep[e.g.,][]{Hayashi:06,Farrar:06,Milosavljevic:07,Springel:07}.
Typically the shock Mach number is inferred from the compression
ratio $r$, which can be measured relatively accurately in X-ray
maps, under the strict assumption that all pressures are thermal
(the observed constraints on $\tau$ are typically used for
consistency check with this assumption). Taking the nonthermal
pressure into account (but ignoring cosmic ray escape; see
\S~\ref{sec:escape}), the Mach number is given by
\begin{equation}\label{eq:mach}
{\cal M}^2=\frac{2r\gamma_{\rm u}/(\gamma_{\rm u}-1)-2\gamma_{\rm
d}/(\gamma_{\rm d}-1)}{\gamma_{{\rm g,u}}(1-\epsilon_{{\rm nt,
u}})(1-r^{-1})[2\gamma_{\rm d}/(\gamma_{\rm d}-1)-r-1]},
\end{equation}
where $\gamma_{\rm g,u} = 5/3$ is the adiabatic index of the
upstream thermal gas, for which relativistic corrections are
negligible. This relation implies that high nonthermal pressure in
the upstream (downstream) increases (decreases) ${\cal M}$. For
example, assuming that $\epsilon_{\rm nt, u}=0$ and cosmic rays are
rather efficiently accelerated in the shock, $\epsilon_{\rm
acc}=0.15$, the inferred value of the Mach number must be revised by
$\approx 10\%-20\%$ downward of the value inferred for a thermal
shock, when $r$ is in the range $2-3$.

If, in addition to $r$, the value of $\tau$ is accurately measured,
the constraints that can be placed on the Mach number are tighter,
since the upstream and downstream nonthermal components are no longer
entirely free. In this case, when $r$ and $\tau$ are held constant, a
larger nonthermal pressure results in a higher Mach number. For
example if the nonthermal component behaves as relativistic gas
($\gamma_{\rm nt}=4/3$) and $r$ and $\tau$ are related as they would
be in a purely thermal shock, $\tau=(4r-1)/r(4-r)$, the shock Mach
number equals
\begin{equation}\label{eq:mach_thermal}
    {\cal M}^2=\frac{3(4r-1)(1-\epsilon_{\rm nt,u})/(4-r)+5r+3r\epsilon_{\rm nt,
    u}-8}{\gamma_{\rm u}(1-\epsilon_{\rm nt,u})(1-r^{-1})(7-r)} ,
\end{equation}
in which case ${\cal M}$ increases by $\approx 10\%$ for
$\epsilon_{\rm nt,u}=0.3$, compared to the purely thermal shock, for
$r$ in the range $2-3$.

\subsection{Turbulence}
\label{sec:turbulence}

In addition to the cosmic rays and the magnetic field, turbulence
also contributes pressure to the ICM. Turbulence in the ICM is
expected to be driven by gravitational clustering (accretion and
merging), and by outflows associated with active galactic nuclei.
Hydrodynamical simulations of galaxy cluster formation in which the
ICM is treated as an ideal fluid universally demonstrate that
turbulent pressure in the ICM is non-negligible, and that its
contribution to the total pressure is an increasing function of
radius from the center of the cluster. The fractional turbulent
pressure measured in the simulations is $\epsilon_{\rm tur}\sim
0.06-0.36$ \citep{Norman:99}, $\epsilon_{\rm tur}\sim 0.05-0.1$
\citep{Ricker:01}, $\epsilon_{\rm tur}\sim 0.04-0.09$
\citep{Nagai:03,Faltenbacher:05}, and $\epsilon_{\rm tur} \sim
0.05-0.3$ \citep{Dolag:05}.\footnote{Hydrodynamic simulations of
intracluster turbulence were also recently carried out by \citet
{Fujita:04}, \citet{Subramanian:06}, and \citet{Vazza:06}.}
Spatially-resolved gas pressure maps obtained from {\it XMM-Newton}
observations of the Coma galaxy cluster show a scale-invariant
pressure fluctuation spectrum on the scales of $40$ to $90\textrm{
kpc}$, which was analyzed to place a lower limit on the fractional
turbulent pressure of $\epsilon_ {\rm tur}\gtrsim 0.1$
\citep{Schuecker:04}.\footnote{Turbulence in the ICM can be detected
in other clusters given an X-ray detector with high spectral
resolution \citep{Inogamov:03,Sunyaev:03}.}

The physics of the interaction of a shock wave with a turbulent
upstream is complex and poorly understood even in unmagnetized,
ideal fluids. The primary theoretical uncertainties are the amount of
amplification of turbulence in the shock and the affect of
turbulence on the shock structure. The shock transition becomes
nonplanar in the presence of turbulence \citep[e.g.,][]{Rotman:91},
and this nonplanarity affects the local shock jump conditions that
short-wavelength fluctuations experience while crossing the shock
\citep[e.g.,][]{Zank:02}. Different analytical approximations (e.g.,
the rapid distortion theory and the local interaction analysis) and
direct numerical simulations do not always agree with each other
\citep [e.g.,][and references therein]{Andreopoulos:00}. Therefore
we do not attempt to include turbulent pressure in our quantitative
calculations. However, since turbulent pressure can affect the
observed jump conditions, we discuss it qualitatively using a simple
model \citep{Lele:92} based on the rapid distortion theory applied
to homogeneous turbulence \citep[see, e.g.,][]{Batchelor:53,Jacquin:93}.

\citet{Lele:92} derives the averaged density, momentum and energy
conservation equations of ideal fluid in the shock frame (his
equations $9-11$). Assuming homogeneous turbulence and a
cylindrically symmetric distribution of turbulent fluctuations, the
conservation equations are reduced to the form of equations
(\ref{EQ:shock_jump}) with a turbulent pressure and effective
adiabatic index of
\begin{eqnarray}
P_{{\rm tur},i} = \overline{\rho_i}\widetilde{v''_{\|,i}
v''_{\|,i}} , \ \ \ \
%\\\nonumber
\gamma_{\rm tur} = \frac{3+2b_i}{1+2b_i} .
\end{eqnarray}
Here, the notation is such that any fluctuating quantity $f$ is
decomposed in two ways $f=\overline{f}+f'=\widetilde{f}+f''$, with
$\overline{f}$ denoting the average value of the quantity,
$\widetilde{f}\equiv \overline{\rho f}/\overline{\rho}$ denoting the
mass-weighted average, and $f'$ and $f''$ denoting the corresponding
fluctuating parts, while as before, $i=({\rm u},{\rm d})$. Just as we did for
the magnetic field (see \S~\ref{sec:magnetic_field}), we use
cylindrical symmetry to parameterize the turbulent field with a
single parameter $b_i=\widetilde{v''_{\perp,i}
v''_{\perp,i}}/(2\widetilde{v''_{\|,i} v''_{\|,i}})$, which in the
isotropic case equals unity.

The temperature jump $\tau$ and the shock Mach number ${\cal M}$ can
be calculated for a given upstream turbulent pressure fraction
$\epsilon_{\rm tur,u} \equiv P_{\rm tur,u}/P_{\rm u}$ and anisotropy
parameter $b_{\rm u}$ if the amplification of the turbulence in the
shock is known.\footnote{In case of isotropic turbulence
$\epsilon_{\rm tur}=\gamma_{\rm g}{\cal M}_{\rm
tur}^2/(3+\gamma_{\rm g}{\cal M}_{\rm tur}^2)$, where ${\cal M}_{\rm
tur}$ is the turbulence Mach number and $\gamma_{\rm g}$ is the
thermal gas adiabatic index.} \cite{Lele:92} uses the rapid
distortion theory to derive the shock amplification. This
approximation assumes that the mean turbulent flux amplitudes are
much smaller than their mean flow counterparts, that turbulent
fluctuations cross the shock much faster than the corresponding eddy
turnaround times, and that the mean flow does not vary much on the
length scale of an eddy. In particular, the otherwise planar shock
transition is assumed to have not been distorted, and rendered
nonplanar, by the fluctuations. In this theory, the parallel and
perpendicular turbulence is amplified according to \citep{Lele:92}
\begin{eqnarray}
\frac{\widetilde{v''_\| v''_\|}_{\rm u}}{\widetilde{v''_\|
v''_\|}_{\rm d}} &=&
\frac{3}{4}r^2\left[\frac{1}{r^2-1}+\frac{r^2-2}{(r^2-1)^{3/2}}\tan^{-1}\sqrt{r^2-1}\right]\nonumber\\&\approx& \frac{6r-1}{5},\nonumber \\
\frac{\widetilde{v''_\perp v''_\perp}_{\rm u}}{\widetilde{v''_\perp
v''_\perp}_{\rm d}} &=&
\frac{3}{8}\left[1-\frac{1}{r^2-1}+\frac{r^4}{(r^2-1)^{3/2}}\tan^{-1}\sqrt{r^2-1}\right]
\nonumber\\&\approx& \frac{r+1}{2},
\end{eqnarray}
which in the case of isotropic upstream turbulence, $b_{\rm u}=1$
and $\gamma_{\rm tur,u}=5/3$, yields effective adiabatic index
and downstream pressure fraction for turbulence
\begin{eqnarray}
\gamma_{\rm tur,d} &\approx& \frac{23r+2}{11r+4} ,\\\nonumber %~~; ~~\left(b_{\rm d} \approx \frac{5(r+1)}{2(6r-1)}\right)\\\nonumber
\epsilon_{\rm tur,d} &\approx& \frac{\epsilon_{\rm
tur,u}(6r-1)}{5\tau(1-\epsilon_{\rm tur,u})+\epsilon_{\rm
tur,u}(6r-1)} .
\end{eqnarray}

In this simple model of amplification of turbulence in the shock,
the effect of a pre-existing turbulent component is similar to that
of the pre-existing cosmic rays and magnetic pressure, namely, the
upstream turbulence is only weakly, positively correlated with
$\tau$ and ${\cal M}$. For example, assuming that turbulence is the
only nonthermal component, the temperature jump increases by a
factor $\approx 1.15$ and ${\cal M}$ increases by a factor $\approx
1.3$ between $\epsilon_{\rm tur,u}=0$ and $\epsilon_{\rm tur,u}=0.3$
for compression ratio $r=2.3$. Very similar results are obtained if
instead of the rapid distortion theory, we employ the linear
interaction analysis to calculate the jump conditions
\citep[e.g.,][]{Lee:93,Lee:97}. Therefore, we conclude that for a
weak, subsonic, unmagnetized turbulence of the ICM, the effect of
turbulence on the thermal gas temperature jump of the shock is
similar to that expected in the presence of other nonthermal
components.

\section{Results}
\label{sec:results}
\subsection{A520}

\cite{Markevitch:05} analyze a $67$ kilosecond observation with {\it
Chandra} ACIS-I of the bow shock in the galaxy cluster merger A520
at z=0.203 and estimate the density and temperature jump along the
axis of symmetry of the shock. The upstream and downstream
temperatures are, respectively, $T_{\rm u}=4.8_{-0.8}^{+1.2}\textrm{
keV}$ and $T_{\rm d}=11.5_{-3.1}^{+6.7}\textrm{ keV}$, while the
density jump is $r=2.3 \pm 0.3$ (all errors are at the $90\%$
confidence levels). These values are consistent with a shock Mach
number of ${\cal M}\approx 2$. We use equation (\ref{EQ:Tjump}) to
explore the constraints that can be placed on the acceleration of
particles in the shock and on the fractional pressure in the
nonthermal components in the shock upstream, and equation
(\ref{EQ:Tjump_gen}) to constrain energy leakage from the shock in
A520.

The predicted downstream temperature in the case of a pure thermal
gas, $T_{\rm d}=10\textrm{ keV}$ for $T_{\rm u}=4.8\textrm{ keV}$
and $r=2.3$, is consistent with the measured value. However, given
that the expected temperature is below the median measured
temperature, little room is left for significant particle
acceleration in the shock. To place constraints on $\epsilon_{\rm
acc}$, we first take $\epsilon_{B,i}=Q=0$, and carry a Monte Carlo
search in the remaining parameter space $(\epsilon_{\rm
CR,u},\epsilon_{\rm acc})$. We draw a large set ($10^5$) of the
observed parameters (density and temperature jump) from the observed
distributions.\footnote{We approximate each observed distribution by
two half-gaussians that peak at the median observed value and
satisfy the $90\%$ confidence range reported by
\citet{Markevitch:05}.} For each set of observed values we find all
the combinations of $(\epsilon_{\rm CR,u},\epsilon_{\rm acc})$ in
the domain $(0<\epsilon_{\rm CR,u}<0.3,0<\epsilon_{\rm acc}<0.25)$
that are compatible with the observations. For each point in the
$(\epsilon_{\rm CR,u},\epsilon_{\rm acc})$ plane, we calculate the
number of instances that the corresponding shock is compatible with
the generated ``observed'' values of $r$ and $\tau$.  The resulting
number, properly normalized, provides the Bayesian likelihood of the
shock being characterized by a given pair $(\epsilon_{\rm
CR,u},\epsilon_{\rm acc})$, assuming a uniform prior in the domain
considered here.

\begin{figure}
\includegraphics[width=9cm]{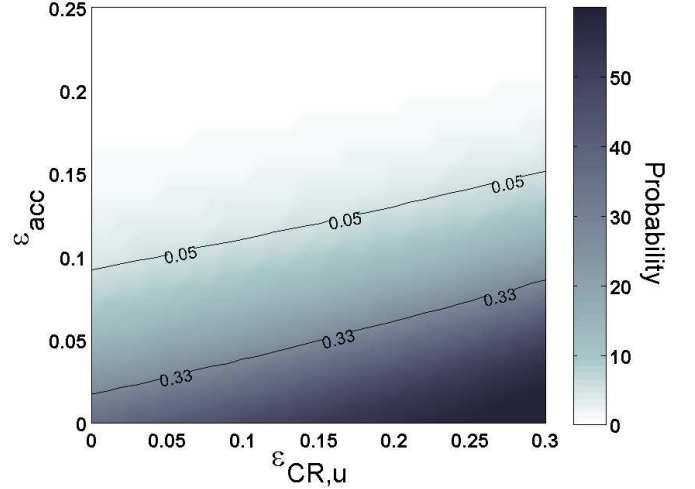}
\caption{ \label{FIG A520} The probability distribution of
$(\epsilon_{\rm CR,u}$,$\epsilon_{\rm acc})$ for relativistic cosmic
rays in the bow shock in the galaxy cluster A520, assuming
$\epsilon_{B,{\rm u}}=0$ and $\epsilon_{\rm CR,u}<0.3$. The contours
divide the plane so that the cumulative distribution above the
contour includes only 0.33 ({\it lower contour}) and 0.05 ({\it
upper contour}) of the total probability. The probability is
calculated via a Monte Carlo simulation (see text).}
\end{figure}

\begin{figure}
\includegraphics[width=9cm]{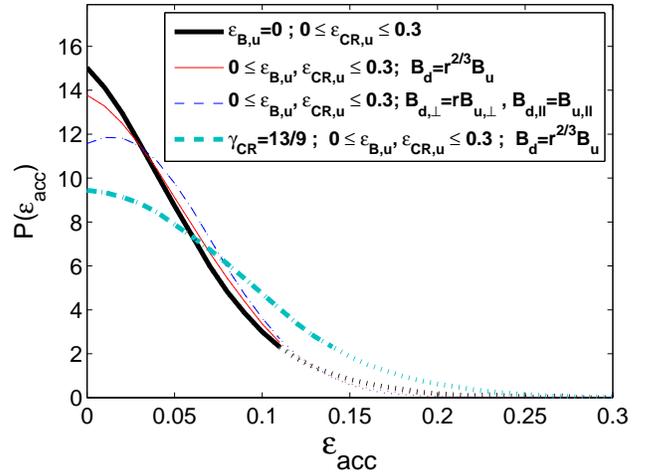}
\caption{ \label{Fig eps_acc_A520} The marginalized probability
distribution of $\epsilon_{\rm acc}$ for several different
nonthermal contributions to the gas pressure. The probability is
calculated via a Monte Carlo simulation (see text) for various
ranges of flat priors on $\epsilon_{\rm CR,u}$ and $\epsilon_{B,{\rm
u}}$, and various levels of magnetic field amplification in the
shocks, as indicated in the legend. The cosmic rays are relativistic
($\gamma_{\rm CR}=4/3$), unless noted otherwise. The solid and
dashed lines contain $90\%$ of the distributions (the rest is in
dotted line tail). This figure shows that the limits on
$\epsilon_{\rm acc}$ are insensitive to the assumptions and the
priors that we choose for the upstream nonthermal components.}
\end{figure}

Figure \ref{FIG A520} shows the probability distribution for
$(\epsilon_{\rm CR,u},\epsilon_{\rm acc})$ for the relativistic
cosmic rays ($\gamma_{\rm CR}=4/3$). The two contours divide the
plane so that the cumulative probability constrained above each is
$0.33$ and $0.05$ of the total. The figure shows that $\epsilon_{\rm
acc} \lesssim 0.1$ at $95\%$ confidence levels for any value of
$\epsilon_{\rm CR,u}$, and that current observations do not provide
a constraint on $\epsilon_{\rm CR,u}$. Carrying out a similar
analysis for relativistic electrons and Newtonian proton cosmic rays
($\gamma_{\rm CR}=13/9$), which would be expected if the cosmic ray
pressure were dominated by particles with typical energies of
$10^1-10^4$ times the thermal energy, yields an upper limit of
$\epsilon_{\rm acc} \lesssim 0.15$. Finally, the limit on the
efficiency of the acceleration of Newtonian cosmic rays,
$\gamma_{\rm CR}=5/3$, as expected if the cosmic rays are
accelerated only to several times the thermal temperature, is
$\epsilon_{\rm acc} \lesssim 0.25$. Naturally, allowing for $Q>0$
when deriving the constraints on $\epsilon_{\rm acc}$ yields a
tighter upper limit on $\epsilon_{\rm acc}$. The minor effect of
varying upstream magnetic field and upstream cosmic ray pressure on
the limits that can be placed on $\epsilon_{\rm acc}$ is explored in
Figure \ref{Fig eps_acc_A520}.

Repeating the same analysis for various values of $0<\epsilon_{B,{\rm
u}}<0.3$ while assuming $\epsilon_{\rm CR,u}=0$, or taking a constant
$\epsilon_{\rm CR,u}$ and a flat prior on $\epsilon_{B,{\rm u}}$ in
the same range, for the two limiting forms magnetic field behavior at the shock
($B_{\bot,{\rm d}}=rB_{\bot,{\rm u}}$ and $B_{\rm d}=r^{2/3}B_{\rm
u}$; see \S~\ref{sec:magnetic_field}), yields a similar upper
limit on $\epsilon_{\rm acc}$. The reason for the weak impact of
magnetic pressure on the constraints that can be placed
on particle acceleration is the
strong dependence of $\tau$ on $\epsilon_{\rm acc}$ and its weak
dependence on $\epsilon_{B,{\rm u}}$. Therefore, just as we found for
$\epsilon_{\rm CR,u}$, current observations do not provide a significant
constraint on $\epsilon_{B,{\rm u}}$.

We also carried out the same analysis assuming no particle
acceleration, $\epsilon_{\rm acc}=0$, but allowing for an escape of
cosmic rays from the shock, in order to constrain energy leakage
from the shock, $Q$. Although this scenario is artificial (since for
$\epsilon_{\rm acc}=0$ we also expect $Q=0$), this analysis provides
an upper limit to the value of $Q$. We find that the observation of
A520 limit $Q\lesssim 0.1$ in its merger shock.

The dependence of the temperature jump in equation (\ref{EQ:Tjump})
on $\epsilon_{\rm CR,u}$ and $\epsilon_{B,{\rm u}}$ is weak.
Therefore current observations do not put significant constraint the
presence of a relativistic component in the pre-merger ICM. Figure
\ref{FIG A520} shows that a future, improved X-ray spectroscopy
across the bow shock, such as with a longer exposure with {\it
Chandra} or {\it XMM-Newton}, may exclude the purely thermal
scenario, $\epsilon_{\rm acc}=\epsilon_{\rm CR,u}=0$. If measured
value of $\tau$ falls below the thermal prediction, our analysis
will yield a lower limit on $\epsilon_{\rm acc}$ to accompany the
current upper limit. If the measured value of $\tau$ falls above the
thermal prediction, the measurement will imply a lower limit on
$\epsilon_{\rm CR,u}$. For example, if we artificially reduce the
present measurement uncertainties in $T_{\rm u}$, $T_{\rm d}$, and
$r$ by a factor of $3$ while assuming that the mean values of these
observables remained unchanged, the data would require
$\epsilon_{\rm acc}<0.05$ and $\epsilon_{\rm nt,u}>0.05$ at $95\%$
confidence levels. However decoupling $\epsilon_{\rm nt,u}$ into its
cosmic ray and magnetic field components cannot be accomplished
given a measurement of $r$ and $\tau$ alone.

From the density jump data alone and assuming a purely thermal
shock, \cite{Markevitch:05} estimate the Mach number of the merger
shock in A520 to be ${\cal M} =2.1_{-0.3}^{+0.4}$. As we discuss in
\S~\ref{sec:mach_number}, the likely presence of a non-negligible
nonthermal pressure requires a modification of the Mach number
estimate as in equation (\ref{eq:mach}); adding a nonthermal
pressure in the upstream increase the Mach number. Figure \ref{Fig
Mach} shows the Mach number probability for several scenarios in
which a nonthermal fluid is present with a fractional contribution
to the pressure that is allowed by the present observations. The
true value of the Mach number depends on the fractional nonthermal
pressure and the amplification of the magnetic field and turbulence
in the shock. For example, for $\epsilon_{\rm CR,u}=\epsilon_{B,{\rm
u}}=0.15$ and assuming significant magnetic field amplification in
the shock, the true Mach number can be as high as ${\cal M} \approx
2.7$

\subsection{1E 0657$-$56}

Gas properties across the merger shock in 1E 0657$-$56 (the
``bullet'' cluster at z=0.296) were measured by \cite{Markevitch:06}
using a $500$ kilosecond observation with {\it Chandra} ACIS-I. They
find a density jump of $r \approx 3$, which corresponds to a Mach
number of ${\cal M} \approx 3$.  The measured temperatures are
$T_{\rm u} \approx 9\textrm{ keV}$ and a lower limit $T_{\rm
d}>32\textrm{ keV}$ at $1\sigma$ confidence levels. The shock in 1E
0657$-$56 is stronger than that in A520 and is thus more propitious
for detecting particle acceleration. Unfortunately, the high
downstream temperature $T_{\rm d}$ complicates an accurate
measurement of the temperature jump with the high resolution X-ray
telescopes ${\it Chandra}$ and ${\it XMM-Newton}$.

\cite{Markevitch:06} does not report the errors on some of the
measurements and therefore we cannot quantitatively constrain the
presence of nonthermal components in the shock upstream and
downstream. Figure \ref{FIG bullet} shows the model prediction for
$T_{\rm d}$ as a function of $\epsilon_{\rm CR,u}$ and
$\epsilon_{\rm acc}$ for the case of relativistic cosmic rays,
$\gamma_{\rm CR}=4/3$, assuming that $\epsilon_{B,{\rm u}}=Q=0$ and
taking $r=3$ and $T_{\rm u}=9\textrm{ keV}$. The figure shows that
constraints that can be made in 1E 0657$-$56 are qualitatively
similar to, although less stringent than, those in A520. The minimum
value of the temperature jump $\tau$ allowed by the measurement is
high and barely consistent with a purely thermal shock.  It does not
leave much room for particle acceleration; we tentatively infer
$\epsilon_{\rm acc}<0.15$.

The nonthermal components may affect also the Mach number of the
bullet cluster merger shock, which recently stirred a discussion about
its compatibility with standard cosmological models
\citep{Hayashi:06,Farrar:06,Milosavljevic:07,Springel:07}.
\cite{Markevitch:06} finds ${\cal M}=3.0 \pm 0.4$ , which corresponds to
$r=3.0^{+0.17}_{-0.23}$, assuming no nonthermal contribution and
neglecting relativistic corrections due to the high electron
temperature. Allowing for cosmic rays with pressure up to
equipartition, $0<\epsilon_{\rm CR,u}<0.3$ and $\epsilon_B=0$, while
requiring temperature  and density jump
consistent with observations, $T_{\rm d}> 20\textrm{ keV}$
and $2.77<r<3.17$, yields the limits
$2.3<{\cal M}<3.7$. The lower bound
is obtained for $r=2.77$, $\epsilon_{\rm acc}=0.07$,
$\epsilon_{\rm CR,u}=0$, and $T_{\rm d}=20\textrm{ keV}$.  The upper bound
corresponds to $r=3.17$, $\epsilon_{\rm acc}=0$, $\epsilon_{\rm
CR,u}=0.3$, and $T_{\rm d}=45\textrm{ keV}$.

\begin{figure}
\includegraphics[width=9cm]{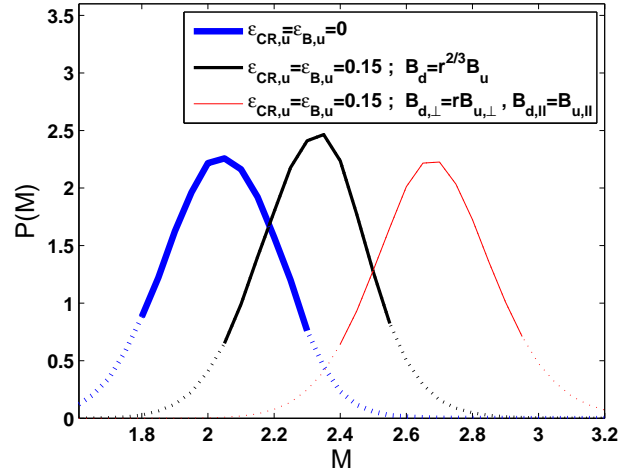}
\caption{ \label{Fig Mach} The Mach number probability distribution
based on the observed density and temperature jumps across the shock
in A520 for several different contributions of nonthermal components
to the gas pressure. The solid lines contain $90\%$ of the
distributions (the rest is in dotted line tails). In all cases the
cosmic rays are relativistic and the upstream magnetic field is
isotropic ($\gamma_{\rm CR}=\gamma_{B,{\rm u}}=4/3)$. The legend
indicates the fractional upstream nonthermal pressure and the shock
amplification of the field in each case.}
\end{figure}

\begin{figure}
\includegraphics[width=9cm]{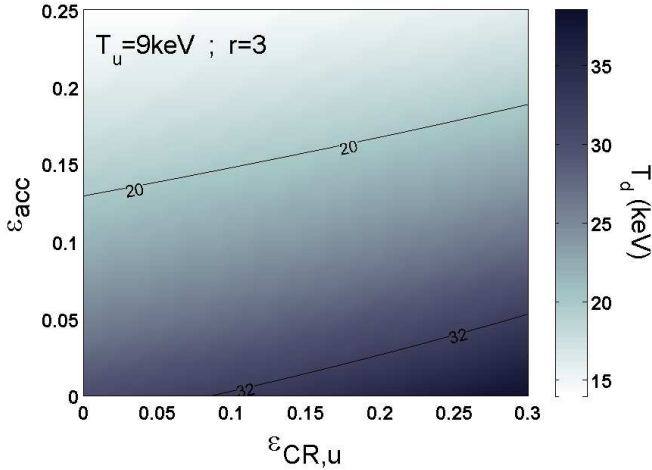}
\caption{ \label{FIG bullet} $T_{\rm d}$ as a function of
$\epsilon_{\rm CR,u}$ and $\epsilon_{\rm acc}$ ($\gamma_{\rm
CR}=4/3$) for $r=3$ and $T_{\rm d}=9\textrm{ keV}$, which correspond
to the values reported for the bullet cluster 1E 0657$-$56. The
$32\textrm{ keV}$ and $20 \textrm{ keV}$ contours are the $1\sigma$
and $2\sigma$ lower limits on $T_{\rm d}$ as measured by
\cite{Markevitch:06} (the $2\sigma$ limit is a rough estimate
derived from the plots in \citealt{Markevitch:06}).}
\end{figure}

\section{Discussion}
\label{sec:discussion}

\subsection{Particle Acceleration in Collisionless Shocks}
\label{sec:accel}

Current measurements place constraints on the efficiency of particle
acceleration in cluster merger shocks.  This offers a very unique
opportunity to test models of particle acceleration in astrophysical
collisionless shocks.  Estimating the acceleration efficiency in
collisionless shocks is a difficult problem, which is severely
complicated by the fact that it remains unknown which among a number
of possibilities is the primary acceleration mechanism. Even in the
leading candidate mechanism, the diffusive shock acceleration,
estimates of acceleration efficiency range widely because of a
number fundamental theoretical uncertainties, concerning the
fraction of thermal particles that are injected into the
acceleration process \citep[e.g.,][and references
therein]{Malkov:01}, the nonlinear influence of the accelerated
particles on the hydrodynamic profile of the shock wave
\citep[e.g.,][]{Drury:81,Achterberg:84,Giacalone:97,Kang:02,Kang:07a},
and the amplification of the magnetic field by plasma instabilities
\citep
[e.g.,][]{Lucek:00,Bell:01,Schlickeiser:03,Bell:04,Bell:05,Schekochihin:05,Medvedev:06}.
Therefore, any estimate of particle acceleration efficiency is
strongly affected by the specific assumptions and approximations
employed in a self-consistent shock model. Here we do not attempt to
carry out a detailed comparison of our results with various
scenarios for particle acceleration.  But, to demonstrate the power
of the constraints that can be obtained from merger shock dynamics,
we discuss our results in the context of the predictions of the
particle acceleration model of \citet[][see also
\citealt{Kang:07b}]{Kang:07a}.

A variety of investigations point to a strong dependence of
acceleration efficiency on the shock Mach number, whereby  stronger
shocks produce higher efficiencies. \citet{Kang:07a} report an
investigation of acceleration efficiency behavior in diffusive shock
acceleration simulations in quasi-parallel shocks with a Bohm
diffusion coefficient, a self-consistent treatments of particle
injection from the thermal pool into the acceleration process, and
Alfv\'en wave propagation. In their Figure 5, \citet{Kang:07a} plot
the dependence of the acceleration efficiency $\eta ({\cal M})$ on
the shock Mach number ${\cal M}$, where $\eta$ is defined as the
energy flux in downstream cosmic rays, divided by the bulk kinetic
energy flux in the upstream medium entering the shock. In the limit
$\epsilon_{\rm acc}\ll1$, the parameter $\eta$ in Kang et al. is
related to our $\epsilon_{\rm acc}$ via
\begin{equation}
\eta \approx \frac{3}{10} \frac{({\cal M}^2+3)(5{\cal M}^2-1)}{{\cal
M}^4} \epsilon_{\rm acc}\ \ \ \ (\epsilon_{\rm acc} \ll 1) .
\end{equation}
Thus, we have $\eta\approx 2.5\epsilon_{\rm acc}$ for ${\cal M}=2$ and
$\eta\approx 2\epsilon_{\rm acc}$ for ${\cal M}=3$.\footnote{We have
here ignored the modification of shock Mach number by the nonthermal
pressure, see \S~\ref{sec:mach_number}.} \citet{Kang:07a} detect a
strong dependence of $\eta$ on the presence of preexisting cosmic rays
in the shock upstream, $\epsilon_{\rm CR,u}$. The strong dependence
can be attributed to inefficient injection at low Mach numbers in
their model. For ${\cal M}=2$, the parameter $\eta$ jumps from $0$ to
$0.15$ as $\epsilon_{\rm CR,u}$ increases from $0$ to $0.23$. In
stronger shocks, for ${\cal M}=3$, the parameter $\eta $ jumps from
$0.1$ to $0.25$ for the same increase in $ \epsilon_{\rm CR,u}$.

Converting our current limits on $\epsilon_{\rm acc}$ into limits on
$\eta$, we find $\eta \lesssim 0.2$ for ${\cal M} \approx 2$ and $\eta
\lesssim 0.3$ for ${\cal M} \approx 3$. These limits are marginally
consistent with the predictions of \citet{Kang:07a} for any assumed
value of $\epsilon_{\rm CR,u}$. However, if as \citet{Kang:07a} argue,
$\epsilon_{\rm acc}$ is itself a sensitive function of $\epsilon_{\rm
CR,u}$, then improved measurements of $\epsilon_{\rm acc}$ that can be
obtained with additional observations with existing X-ray telescopes
can provide tighter limits on $\epsilon_{\rm CR,u}$. Moreover, a
positive measurement of $\epsilon_{\rm acc}$ can provide a
model-dependent constraint on $\epsilon_{\rm CR,u}$.

\subsection{X-ray and SZE Cluster Mass Estimates}
\label{sec:mass}

Galaxy cluster surveys in which the cluster masses are measured
accurately can be used as powerful cosmological probes of dark matter
and dark energy.  The mass estimates are plagued by systematic
uncertainties that must be understood and quantified before the
requisite mass measurement accuracy is achieved.  Nonthermal pressure
due to cosmic rays, magnetic fields, and turbulence, is a source of a
systematic bias when cluster masses are estimated on the basis of the
assumption of hydrostatic equilibrium between gravitational forces and
thermal pressure gradients in the ICM \citep*[e.g.,][and references
therein]{Ostriker:05,Rasia:06,Nagai:07a}.  The hydrostatic mass
profile of a spherically-symmetric cluster is given by
\begin{equation}
 M(<r) = \frac{-r^2}{G\rho_{\rm g}} \left( \frac{dP_{\rm
 g}}{dr}+\frac{dP_{\rm nt}}{dr} \right),
\label{eq:HSE}
\end{equation}
where $M(<r)$ is the mass enclosed within radius $r$, while $P_{\rm
g}$ and $P_{\rm nt}$ are the thermal and the nonthermal
contributions to the pressure.  The thermal gas provides a
significant fraction of the total pressure support, and this
pressure is measured directly with current X-ray and SZE
observations.  The contribution of the nonthermal pressure, on the
other hand, is customarily assumed to be relatively small ($\lesssim
10\%$) outside of a cluster core \citep[see e.g.,][]{Nagai:07b}, and
it is often ignored in the hydrostatic mass estimates based on X-ray
and SZE data. However, present observations do not yet constrain the
nonthermal pressure in the regime in which it dramatically affects
the calibration of the hydrostatic mass estimates. If not accounted
for, these nonthermal biases limit the effectiveness of upcoming
X-ray and SZE cluster surveys to accurately measure the expansion
history of the universe. Detailed investigations of the sources of
nonthermal pressure in clusters are thus critical for understanding
their effect on the properties of the ICM and the utility of
clusters as precision cosmological probes. We proceed to discuss how
future observations of cluster merger shocks will have the potential
to place unique constraints on the nonthermal pressure in the
unshocked ICM, thereby improving cluster mass estimates.

\subsection{Prospects for Future  Constraints of Nonthermal Pressure}
\label{sec:future}

As discussed in \S\ref{sec:results}, current measurements alone do
not place strong constraints on the presence of a nonthermal
component in the unshocked ICM in both systems.  However, the
improved constraints that can be obtained with existing X-ray
telescopes can provide useful lower limits on nonthermal pressure
and their effects on the X-ray and SZE cluster mass estimates. In
the case of the shock in the cluster merger A520, the current
constraints, which are based on a 67 kilosecond observation with
\emph{Chandra}, can be improved significantly with follow-up
observations with \emph{Chandra} or \emph{XMM-Newton}. Therefore, we
identify this system as the most promising one in which our method
may yield a positive shock-hydrodynamic detection of a nonthermal
component.  While the stronger shock in the cluster merger 1E
0657$-$56 may be more efficient at accelerating particles, it will
be more difficult to improve the measurement of the shock
temperature jump in 1E 0657$-$56. This is because the very high
temperature of the shock downstream medium ($T_{\rm d}\approx
30-50\textrm{ keV}$) lies far outside of spectral sensitivity window
of X-ray telescopes with arcsecond resolution, which in turn renders
it difficult to measure the temperature jump in the narrow
post-shock layer. Hard X-ray observations
\citep[e.g.,][]{Petrosian:06} with {\it RXTE}, \emph{Integral}, or
{\it Suzaku}, combined with a model of ICM fluid flow
\citep[e.g.,][]{Milosavljevic:07,Springel:07}, may help to pin down
the downstream temperature. Alternative intriguing possibility is
that future high resolution SZE observations may be able to measure
the downstream temperature on the basis of the relativistic SZE,
which should be prominent in the high temperature downstream.

In the next few years, gamma-ray observations of galaxy clusters may
provide tight constraints on the fractional contribution of nonthermal
particles to the pressure of the ICM. Assuming that gamma-ray emission
from the decay of neutral pions is the primary emission channel,
measurements with the new gamma-ray telescope \emph{GLAST} can be used
to place population-averaged limits on the hadronic cosmic-ray
pressure support in clusters \citep[][see also
\citealt{Berrington:03,Blasi:07}]{Ando:07}. These forthcoming
constraints, combined with improved X-ray spectroscopic measurements
of the jump conditions in merger shocks, may enable a separation of
the upstream nonthermal pressure into its constituent components.

Finally, comparisons of the hydrostatic mass estimates with those
derived from gravitational lensing observations can, in principle,
provide an important handle on nonthermal biases. Note that this
approach is not practical for individual clusters, because lensing
measures the mass in a projected aperture that cannot be directly
compared to the mass within a sphere of the same radius, to which
the hydrostatic mass is sensitive.  But, it might be possible to
compare different estimators in an average sense, while accounting
for the effects of asphericity of clusters and projection effects
\citep[see \S~5.2 in][]{Nagai:07a}.

\section{Conclusions}
\label{sec:conclusions}

We model the effect of particle acceleration and nonthermal pressure
components on shock jump conditions in nonrelativistic shocks.  We
focus on intermediate Mach number shocks, with Mach numbers in the
range ${\cal M}=2-3$. We apply this to the merger shocks in galaxy
clusters A520 and 1E 0657$-$56 and place the first constraints on the
efficiency of particle acceleration in these shocks. Our main results
are as follows.

1. The temperature jump of the thermal gas in the shock depends
  strongly on the efficiency of shock particle acceleration.
  Efficient acceleration can reduce the temperature jump by more than
  a factor of two for a constant compression ratio in the range
  $r=2-3$.

2. The correct effect of nonthermal pressure in the upstream, such as
fossil cosmic rays, magnetic field, and turbulence, on the shock jump
observed in the thermal gas cannot be derived at this point, because
we lack an understanding of the interaction of these components with
the shock.  However, for a wide range of reasonable assumptions and
analytic approximations, we find that nonthermal pressure in the
unshocked ICM has only a minor effect on the downstream temperature
(at a fixed compression ratio), and that in general, a high upstream
nonthermal pressure increases the temperature jump in the thermal gas.

3. The combination of strong dependence of the temperature jump on
  particle acceleration and weak dependence on upstream nonthermal
  pressure enables derivation of meaningful constraints on the
  efficiency of particle acceleration in cluster merger shocks, even
  with current observations. Future, more accurate X-ray and SZE
  observations of these shocks may yield meaningful constraints on the
  upstream nonthermal pressure as well.

4. Nonthermal pressure and shock particle acceleration can also affect
  the Mach number that is inferred from the observed compression ratio
  $r$ by tens of percent.  When the temperature jump is poorly
  constrained, the Mach number is anticorrelated with efficient
  particle acceleration and is positively correlated with upstream
  nonthermal pressure.

5. In the two observed high contrast galaxy cluster merger shocks,
  A520 and 1E 0657$-$56, we constrain the efficiency of acceleration
  of relativistic particles to be $\epsilon_{\rm acc} \lesssim 0.1$
  and $\epsilon_{\rm acc} \lesssim 0.15$, respectively. We find that
  considerable upstream pressure can increase the Mach number of the
  shock in A520 to reach ${\cal M} \approx 2.7$, much higher than the
  inferred value $2.1$ obtained assuming an absence of nonthermal
  components.  The true Mach number of the shock in 1E 0657$-$56 can
  be in the range $2.3<{\cal M}<3.7$ with the compression ratio of
  $r=3$ allowing for nonthermal pressure components.

\acknowledgements

The research was supported in part by the Sherman Fairchild
Foundation. We would like to thank A. K\"onigl, E. Komatsu,
A. Kravtsov, P. Kumar, Y. Rephaeli, and J. Scalo for helpful
discussions.

\end{document}